\documentclass[aps,prl,twocolumn,groupedaddress]{revtex4}
\usepackage{amssymb,amsfonts,amsmath,graphicx,color,mathptm,times}
\usepackage{mathrsfs,natbib}
\usepackage{amsmath,amssymb, amsfonts, bbm}
\usepackage{graphics,epsfig,color,subfigure}
\usepackage{hyperref}
\usepackage{verbatim}
\usepackage{bm}
\usepackage{enumitem}

\usepackage{eufrak}
\usepackage{natbib}

\newcommand{\be}{\begin{equation}}
\newcommand{\ee}{\end{equation}}

\def\bea{\begin{align}}
\def\ena{\end{align}}

\def\tr{\mbox{tr }}
\def\beqa{\begin{eqnarray}}
\def\enqa{\end{eqnarray}}

\hypersetup{
colorlinks=true, % false: boxed links; true: colored links
linkcolor=red,   % color of internal links (change box color w linkbordercolor)
citecolor=blue,       % color of links to bibliography
filecolor=magenta,    % color of file links
urlcolor=blue         % color of external links
}

\newcommand{\avg}[1]{\left\langle #1\right\rangle}

\begin{document}

\title{Global symmetries, volume independence and continuity}
\author{Tin Sulejmanpasic}
\affiliation{Department of Physics, North Carolina State University, Raleigh, NC, 27695\\
and\\
Philippe Meyer Institute, Physics Department, Ecole Normale Sup\'erieure, PSL Research University, 24 rue Lhomond, F-75231 Paris Cedex 05, France}
\email{tin.sulejmanpasic@gmail.com}

\begin{abstract}
We discuss quantum field theories with global $SU(N)$ and $O(N)$ symmetries for which the temporal direction is compactified on a circle of size $L$ with periodicity of fields up to a global symmetry transformation, i.e. twisted boundary conditions. Such boundary conditions correspond to an insertion of the global symmetry operator in the partition function. We argue that for a special choice of twists most of the excited states get projected out, leaving only either mesonic states or states whose energy scales with $N$. When $N\rightarrow \infty$ all excitations become suppressed at any compact radius and the twisted partition function gets a contribution from the ground-state only, rendering observables independent of the radius of compactification, i.e. volume independent. We explicitly prove that this is indeed the case for the $CP(N-1)$ and $O(N)$ non-linear sigma models in any number of dimensions. We further focus on the two-dimensional $CP(N-1)$ case which is asymptotically free, and demonstrate, unlike its thermal counterpart, the twisted theory has commuting $N\rightarrow\infty,\;L\rightarrow\infty$ limits and does not undergo a second-order phase transition at ``zero-temperature'' discussed by Affleck long ago. At finite $L$ the theory is described by an effective, zero-temperature quantum mechanics with smoothly varying parameters depending on $L$, eliminating the possibility of a phase transition at any $L$, which was conjectured by \"Unsal and Dunne. As $L$ is decreased at fixed and finite $N$ the relevant objects dictating the $\theta$ dependence are quantum kink-instantons, avatars of the small $L$ regime fractional instantons. These considerations, for the first time establishes the idea of adiabatic continuity advocated by \"Unsal et. al. 

%short abstract:

%We discuss quantum field theories with global $SU(N)$ and $O(N)$ symmetries for which temporal direction is compactified on a circle of size $L$ with periodicity of fields up to a global symmetry transformation, i.e. twisted boundary conditions. Such boundary conditions correspond to an insertion of the global symmetry operator in the partition function. We argue in general and prove in particular for $CP(N-1)$ and $O(N)$ nonlinear sigma models that large $N$ volume independence holds. Further we show that the $CP(N-1)$ theory is free from the Affleck phase transition confirming the \"Unsal continuity conjecture.
 \end{abstract}

\maketitle

\noindent{\bf Introduction:}
The question of how a short-distance formulation of a given quantum theory is related to its long-distance physics is, in the most interesting cases, an extremely difficult one to understand in general. Although the problem of relating the UV and the IR physics is often associated with \emph{asymptotically free quantum field theories}, it transcends such theories, and the search for low energy descriptions is a problem in many other non-renormalizable systems, from quantum magnets, Weyl semi-metals and topological insulators to cold atom systems. 

The basic problem in this category is establishing whether the system is gapped or ungapped, and what is the nature of its excitations. These questions are not easily answered in general and for non-asymptotically free theories the answer is sensitive to UV parameters. For this reason asymptotically free theories enjoy a distinguished place: they are theories for which there are no tunable UV parameters. In other words, the only tunable parameters are all-scale parameters (e.g. rank of the symmetry group, $\theta$-angle, etc.) The reason for this is that asymptotic freedom allows a rigorous elimination of the cutoff scale, so that the theory is well defined at arbitrarily short distances. The price to pay for this, however, is that when short distance scales are integrated out, the theory flows into a strong coupling regime, rendering standard perturbative and semi-classical methods unreliable and the IR physics obscured. 

Some access to the theory may still be gained by considering regimes in which a large energy scale saturates the coupling and provides a reliable perturbative and semi-classical expansion for some observables. One of the most successful analysis of QCD and some of its non-perturbative effects is that of the high-temperature regime (see e.g. \cite{Gross:1980br}). However this regime is separated from the low-temperature regime by a confinement/deconfinement (and chiral) phase transition, and is characterized by qualitatively different features. For the $SU(N)$ gauge theory the confinement/deconfinement phase transition is marked by the breaking of the so-called center symmetry, a symmetry which acts on static heavy quark wordlines, or Polyakov loops $\mathcal P\rightarrow z\mathcal P$, where $z\in \mathbb Z_N\subset SU(N)$ is the center element (see \cite{Gaiotto:2014kfa} for a modern discussion). In the confined regime the static heavy quark is confined and so $\avg{\mathcal P}=0$ as it has infinite free energy, while in the deconfined regime $\avg{\mathcal P}\ne0$, so that the $\mathbb Z_N$ symmetry is spontaneously broken.

It was argued long ago by Eguchi and Kawai that a lattice Yang-Mills theory enjoys a large $N$ volume independence, i.e. that in the limit of $N\rightarrow\infty$ the physical observables are insensitive to the size of the space-time volume \cite{Eguchi:1982nm}. However an assumption made by Eguchi and Kawai is that the center symmetry is unbroken for any volume. But as we already mentioned even when only one direction is compactified the confinement/deconfinement phase transition sets in and breaks the center symmetry, rendering EK reduction invalid in the continuum limit \cite{Bhanot:1982sh}. 

Nevertheless some success towards the EK reduction was made by the imposition of certain boundary condition twists to prevent a phase transition in small volume lattice theories, the so called Twisted Eguchi-Kawai (TEK) model \cite{GonzalezArroyo:1982ub,GonzalezArroyo:1982hz}. While the original TEK model fails in the continuum limit \cite{Azeyanagi:2007su}, an improved version of TEK appears to have no such issue \cite{GonzalezArroyo:2010ss}. 

In parallel a program initiated by M. \"Unsal took a different perspective. Either by adding adjoint fermion flavors with non-thermal boundary conditions or by deformation of the action, the stability of the center could be achieved \cite{Unsal:2007jx,Kovtun:2007py,Unsal:2008ch}. Further these works argued continuity in the regime where strict volume independence does not hold, and semi-classics can safely be utilized. This spurred a lot of work in QCD-like theories \cite{Cossu:2009sq,Bringoltz:2009kb,Azeyanagi:2010ne,Unsal:2010qh,Armoni:2011dw,Poppitz:2012sw,Argyres:2012vv,Argyres:2012ka,Basar:2013sza,Misumi:2014raa,Misumi:2014bsa,Anber:2014sda,Anber:2015kea,Cherman:2016hcd,Fujimori:2016ljw,Cherman:2016jtu,Sulejmanpasic:2016uwq} and in various sigma models in two space-time dimensions \cite{Dunne:2012ae,Dunne:2012zk,Cherman:2013yfa,Misumi:2014jua,Dunne:2015ywa}. However in all of the cases the role of twists and the underlying physical cause of volume independence and/or continuity is not entirely clear, and relation to the small and large volume regimes remains obscured and conjectural. 
 
It is the objective of this work to not only clarify the role of volume independence in theories with a global $SU(N)$ (or $O(N)$) symmetry, but also to give the first valid proof of such independence in the concrete example of the $CP(N-1)$ and $O(N)$  sigma models in \emph{any} dimensions. Further we will, for the first time, make explicit contact between large and small circle regimes which are related to the works \cite{Unsal:2007jx,Kovtun:2007py,Unsal:2008ch,Cossu:2009sq,Bringoltz:2009kb,Azeyanagi:2010ne,Unsal:2010qh,Armoni:2011dw,Poppitz:2012sw,Argyres:2012vv,Argyres:2012ka,Basar:2013sza,Misumi:2014raa,Misumi:2014bsa,Anber:2014sda,Anber:2015kea,Cherman:2016hcd,Fujimori:2016ljw,Cherman:2016jtu,Sulejmanpasic:2016uwq}. The importance of making this connection is only matched by the simplicity of our findings. In fact the volume independence in $CP(N-1)$ and $O(N)$ sigma models which we present here is very likely the simplest example of the EK idea.
 
We will see that by imposing twists in the temporal direction so that the elementary fields are periodic up to a global symmetry transformation (i.e. twists) the theory retains a wealth of ground-state properties. The underlying reason is that such twists serve as projectors onto a very restricted Hilbert space. The restricted Hilbert space is dominated by excitations which have masses scaling with $N$. Since such states decouple in the large $N$ limit, they are rendered irrelevant in this limit. We will prove that this is indeed the case in the $CP(N-1)$ and $O(N)$ sigma model in \emph{any space-time dimensions}. 

Although beyond the scope of the present work, we believe our considerations can be generalized to other sigma-models \cite{Cherman:2013yfa,Dunne:2015ywa}. 
%We also comment on the relevance to gauge theories in the conclusions
We also believe that similar explanation exists for the successful implementation of twisted boundary conditions in Yang-Mills theory \cite{GonzalezArroyo:2010ss}. Indeed the twist employed in this work are related to conserved fluxes due to the $\mathbb Z_N$ center symmetry \cite{'tHooft:1979uj}.

\vspace{.25cm}
\noindent{\bf The thermal and the twisted partition functions:} Consider a general system with a global symmetry group $G$. The thermal partition function is given by
\be
Z=\tr e^{-H\beta}\;.
\ee
The path integral formulation of such an object amounts to compactifying time on a circle of size $\beta$. The vacuum properties are most pronounced when the temperature $T=1/\beta$ is small, so that the high-energy states are exponentially suppressed. However it is often the case that the partition function of quantum field theories is easier to compute when $\beta$ is small. The reason for this is that Quantum Field theories are defined by their UV Lagrangian which are typically better controlled then their IR properties due to the large quantum fluctuations at long distances (i.e. flow of the theory to strong coupling). The limit of high temperature is clearly an admixture of all states, and the information about the IR properties of the theory (i.e. the ground state and the mass-gap) are obscured in this limit.

For this reason it is useful to introduce an object similar to the thermal partition function
\be\label{eq:twist_part}
Z_{\Omega}=\tr e^{-HL}\hat\Omega\;.
\ee
where $\hat\Omega$ is a unitary operator implementing \emph{a global symmetry transformation}. In the path-integral formulation, the insertion of $\hat\Omega$ is equivalent to the twisting of the temporal boundary conditions. To distinguish it from the thermal compacitification we use $L$ to denote the size of the compact time.
Because of the insertion $\hat\Omega$, some states belonging to the same irreducible representation of the group $G$ may be weighted by phases (or signs) in the partition function which can cause cancellations between the different states in that representation. It is this observation that is at the heart of the large $N$ volume independence we discuss here. 

By choosing the twist appropriately it is possible to eliminate a large number of excited states completely, which would allow the parameter $L$ to be taken small while still keeping a significant grasp on the ground state properties of the theory. In particular, if the ground state is a singlet under the symmetry transformation $\hat\Omega$ (i.e. there is no spontaneous symmetry breaking \footnote{It is possible this assumption can be relaxed. See \cite{Cherman:2016hcd}.}), it will not receive phase contributions from the insertion of $\hat\Omega$ in the twisted partition function $Z_{\Omega}$. The trick is than to choose $\hat\Omega$ in such a way so that its trace vanishes in ``most'' representations of the group $G$, eliminating a vast number of excited states. 

For concreteness let us consider the symmetry group $SU(N)$. If we choose twists $\Omega$ such that 
\be\label{eq:Omega_fund}
\tr_{F}\Omega^{i}=0\;,\qquad i=1,\dots N-1
\ee
where the trace is in the fundamental representation, then the trace over $\Omega$ in all representations with $k$ fundamental indices (the $k$-index representation) will be zero as long as $k$ is not a multiple of $N$ \footnote{This can be easily seen by noting that eigenvalues of $\Omega$  in the $k$-index representation are given by $e^{2\pi i (n_1+n_2+\dots n_k)/N}$ where $n_1,n_2,\dots n_k$ are integers. The restriction on the integers $n_i$ depends on the particular representations, but for any representation one can redundantly do a replacement $n_i\rightarrow n_i+m$. The integer $m$ can always be eliminated by relabeling dummy indices $n_i$. We can then redundantly sum over $m$. But this sum vanishes unless $k$ is divisible by $N$.}. This means that unless the state transforms under a representation that is a combination of $N$ fundamental indices, the partition function defined with these twists will project these states out \emph{exactly}.  

So what excitations remain? Let us think about exciting quanta in the fundamental representation, and let us assume that such fundamental quanta have a minimal mass $M$. Then the typical $N$-index representations will have a mass scaling with $N$, so that most of them will be ``heavy'' if $N$ is taken to be large. The exceptions are the adjoint and the singlet representations and representations that can be built from their products. The singlet and the adjoint states will contribute to the partition function as
\be\label{eq:singlet-adj}
e^{-E_{\rm s}L}+\tr_{\rm adj}\Omega e^{-E_{\rm adj}L}=e^{-E_{\rm s}L}- e^{-E_{\rm adj}L}\;.
\ee
where $\Omega$ has properties \eqref{eq:Omega_fund} and $E_{\rm s}$ and $E_{\rm adj}$ are energies of the singlet and the adjoint excitations respectively. Notice that the relative sign is due to the identity $\tr_{\rm adj}\Omega=|\tr_F \Omega|^2-1=-1$.

Now in the $N\rightarrow\infty$, the quanta transforming under the fundamental representation of $SU(N)$ generically have trivial S-matrix elements, so that states with equal number of fundamental and anti-fundamental quanta are degenerate. In particular $E_{\rm s}=E_{\rm adj}$ in \eqref{eq:singlet-adj}, and the singlet and adjoint contributions will exactly cancel. Notice however that because of the relative sign in \eqref{eq:singlet-adj} even when $N$ is finite and $E_{\rm adj}-E_{\rm s}\sim \Delta/N+o(1/N^2)$, by taking $L\Delta/N\ll 1$ we are able to suppress contribution \eqref{eq:singlet-adj} from the partition function, leaving only exponentially suppressed states $\sim e^{-MNL}$.  In other words volume independence holds even if $N$ is finite and large as long as we have $\frac{1}{NM}\ll L\ll \frac{N}{\Delta}$.  We must emphasize we assume that all excitations carry group $G$ indices. If this is not the case taking $L$ small may still cause excitations which are not classified by the group $G$.

This discussion can be translated to the $O(N)$ group. 
The subtle issue in this case is that the cancellation happens between symmetric traceless representations and their respective traces which are degenerate in the large $N$ limit. 
We will however not dwell on this in details as it is not very illuminating. Instead we will demonstrate explicitly the large $N$ volume independence in the cases of the $CP(N-1)$ and $O(N)$ nonlinear-sigma models.

\vspace{.25cm}
\noindent{\bf The CP(N-1) and the O(N) nonlinear sigma models:}
Here we work out the example of the $CP(N-1)$ sigma model in detail, and give, for the first time, an explicit and complete proof of large $N$ volume independence. An almost verbatim derivation holds for the $O(N)$ model, as we comment below. The proof is surprisingly elementary.

To begin with the \mbox{$CP(N-1)$} action in $D$-dimensions
\be\label{eq:action}
S=\frac{N}{f_0}\int d^Dx |D_\mu u|^2\;.
\ee
where $u=(u_1,u_2,\dots,u_N)$ is an \emph{$N$-dimensional complex vector field of unit length,} (i.e. $u^\dagger u=1$), the $D_\mu=\partial_\mu+iA_\mu$ is a covariant derivative and $A_\mu$ is a $U(1)$ (auxiliary) gauge field. The theory enjoys a gauge symmetry $u\rightarrow e^{i\phi(x)}u$, $A_\mu\rightarrow \partial_\mu\phi$ which leaves the Lagrangian invariant. When $D= 2$, the theory can be supplemented by a $\theta$ angle term $i\theta\int d^2x\;\epsilon^{\mu\nu}\partial_\mu A_\nu$. Nevertheless non-renormalizable theories for $D>2$, although sensitive to the UV cutoff are valid effective theories and have great utility in condensed matter systems.

The model \eqref{eq:action} can be solved in the large $N$ limit by decomposing the constraint $u^\dagger u=1$ with the help of a Lagrange multiplier $\lambda(x)$, by introducing the  term $\frac{N}{f_0}\int d^Dx\; \lambda(x) (u^\dagger u-1)$ in the action. The action is now quadratic in the $u$-field which can therefore be integrated out, yielding an effective action for the $\lambda$-field and $A_\mu$.

\be\label{eq:CPNbegin}
S_{eff}[A_\mu,\lambda]=-\tr\log(-D_\mu^2+\lambda)-\frac{N}{f_0}\int d^Dx\; \lambda
\ee
Since we consider the twisted boundary conditions 
\be
u(L)=\Omega u(0)
\ee
where $\Omega\in SU(N)$. Taking the basis for $u$ such that $\Omega=\text{diag}(e^{2\pi i \nu_0},e^{2\pi i\nu_1 },\dots,e^{2\pi i \nu_{N-1}})$, with $\sum_{s=0}^{N-1}\nu_s =0\bmod 2\pi$ we have
\begin{multline}
S_{eff}=-\int d^Dx\Bigg(\sum_{s=0}^{N-1}  \frac{1}{L}\sum_n \int \frac{d^{D-1}k}{(2\pi)^{D-1}}\\\times\log\Big[-(D_0+i\tfrac{2\pi(n+\nu_s)}{L})^2-(D_i+ik_i)^2+\lambda(x)\Big]+\frac{N}{f_0}\lambda(x)\Bigg)\;.
\end{multline}
where $i$-index runs along the non-compact directions.
If we choose twists which obey \eqref{eq:Omega_fund}, we can always choose, up to a $U(1)$ gauge transformation, $\nu_s=\frac{s}{N}$. Noting that
\be
n+\frac{s}{N}=\frac{Nn+s}{N}\in \frac{\mathbb Z}{N}
\ee
we can replace
\be
n+\nu_s=\frac{N n+s}{N}\rightarrow \frac{m}{N}\;, \qquad \sum_{s=0}^{N-1}\sum_{n}\rightarrow \sum_m\;.
\ee
The crucial observation is that the sum over the Matsubara index $n$ and the trace index $s$ is combined into a single sum over $m$.  Now we have
\begin{multline}\label{eq:Seff_inter}
S_{eff}=-N\int d^Dx\Bigg(\frac{1}{NL}\sum_{m}   \int \frac{d^{D-1}k}{(2\pi)^{D-1}}\\\times\log\Big[-\left(D_0+i\tfrac{2\pi m}{NL}\right)^2-(D_{i}+ik_i)^2+\lambda(x)\Big]+\frac{N}{f_0}\lambda(x)\Bigg)\;.
\end{multline}
Notice that taking $N\rightarrow \infty$ we have that the sum
\be
\lim_{N\rightarrow \infty}\frac{1}{NL}\sum_m F\left(\frac{2\pi m}{NL}\right)=\int \frac{dk_0}{2\pi}\;f(k_0)\;,
\ee
so that when $N$ is large, we get
\begin{multline}\label{eq:CPNfinal}
 {S_{eff}}=-N\int d^Dx\Bigg( \int \frac{d^Dk}{(2\pi)^D}\\\times\log\Big[-(D_\mu+ik_\mu)^2+\lambda(x)\Big]+\frac{1}{f_0}\lambda(x)\Bigg)\;.
\end{multline}
The right hand side looks like the action on $\mathbb R^D$. There is one difference however as the above temporal extent is $x_0\in[0,L)$. The fields $A_\mu,\lambda$ are, in other words, periodic with respect to this interval. However, in the limit $N\rightarrow \infty$, these fields can be evaluated at their saddles, as their fluctuations are suppressed as $1/N$. Therefore, by replacing $A_\mu,\lambda$ by their saddle point values on $\mathbb R^D$ $A_\mu=0,\lambda=M^2$  \cite{Witten:1978bc}, where $M$ is the mass of the $u$-field excitations, we have that the integrand above is space-time independent. It follows immediately that the free energy is identical to the one on $\mathbb R^D$. This proves that the simple 't Hooft large $N$ limit in the theory with twisted boundary conditions is equivalent to the large $N$ theory on $\mathbb R^D$.
We make several observations:
\begin{enumerate}[leftmargin=*]
\item The result we derived is valid at $N\rightarrow \infty$ for any dimension $D$. For $D>2$, the theory is non-renormalizable, but is perfectly sensible as a lattice theory. 
%Irregardless of the regularization, the volume dependence holds, as long as twists are implemented. The meaning of the partition function with such twists is given by \eqref{eq:twist_part}. 
\item If $N$ is finite, however, the twisted partition function reminds us of a thermal partition function with temperature $1/(NL)$, as can be seen from \eqref{eq:Seff_inter}. In other words the ``Matsubara frequency'' is $2\pi/(NL)$, and the compactification radius is effectively increased $N$ times. This is consistent with our observation that the only excitations charged under the $U(1)$ gauge group which remain in the twisted theory have masses scaling with $N$. 
If we assume that the system is gapped with the gap $M\ne 0$, then in order to induce an admixture of excited states in the system we must have $NL\sim M^{-1}$. As long as $NL\gtrsim M^{-1}$, the effective ``temperature'' is small and the would-be ``thermal'' excitations are suppressed, which is precisely the expectation in gauge theory \cite{Unsal:2010qh}. 

\item The volume independence also holds for any composition of $SU(N)$ invariant operators $u(x)$ and $u^\dagger(x)$. Consider for example an $SU(N)$-invariant operator 
\be
\avg{u^\dagger(x)u(0)}=\tr_\Omega \frac{1}{-D_\mu^2+\lambda}\rightarrow N\int \frac{d^Dk}{(2\pi)^{D}}\frac{e^{ik_\mu x_\mu}}{k_\mu^2+M^2}
\ee
where the trace above is over the $SU(N)$ indices of the propagator, and the subscript $\Omega$ indicates the twists, and where we replaced $\lambda,A_\mu$ by their saddle point values.

\item The case when $D=2$ is asymptotically free and is special. At large but finite $N$ the gauge fields induce a linear (i.e. confining) attraction between the $u$-quanta \cite{Witten:1978bc}, with the string tension $\propto M^2/N$. In the remainder of this work we will focus entirely on the effect of these fluctuations in $D=2$.

\item The same steps can be repeated for the $O(N)$ nonlinear sigma model with the Lagrangian $\frac{N}{f_0}(\partial_\mu \bm{n})^2$ with the twists $\Omega: n_i\rightarrow \Omega_{ij}n_j$ acting on the $N$-dimensional unit vector $\bm n$ with $\bm n^2=1$ such that $\Omega_{ij}=\delta_{i+1\bmod N,j}$. This cyclically permutes the components of the $\bm n$-vector $n_{i}\rightarrow n_{i+1\bmod N}$. 
After similar manipulations as for the case of $CP(N-1)$ the formulas obtained for the $O(N)$ model are identical to \eqref{eq:CPNbegin} through \eqref{eq:CPNfinal} up to replacement $A_\mu\rightarrow 0$. In fact it is most easily seen in the $O(N)$ case that the theory can even be regulated on the lattice with such twists, and the compact direction can be only a single lattice site long. The $N\rightarrow\infty$ limit completely recovers the infinite volume/continuum limit in the relevant direction. 

\end{enumerate}

\vspace{.25cm}
\noindent{\bf $CP(N-1)$ model in 2D and the absence of the large-$N$ phase transition:}
It was shown long time ago by Affleck  \cite{Affleck:1979gy}  in the two-dimensional $CP(N-1)$ model that the large-$N$ and the zero temperature limits do not commute. This was possible to do by a reliable large $N$ integration over all the Matsubara modes, reducing the theory to an effective quantum mechanical system (i.e. a 1D quantum field theory \footnote{Since we think of time as compactified, the remaining spatial direction must be thought of as ``time'' for the quantum mechanical system.}). The conclusion of Affleck is that at $N\rightarrow\infty$ the system has a phase-transition at zero temperature. The source of this phase-transition can be traced to the observation that high temperature phase is a 1D Coulomb plasma phase, which screens the electric charges by Debye screening. Such excitations typically induce a potential for the time component $A_0$ of the gauge field (which becomes a compact scalar)
\be\label{eq:VA0_therm}
V_{eff}(A_0)\propto -Ne^{-M\beta}\cos(A_0\beta)\;,
\ee
where $M$ is the mass of the elementary excitations charged under the gauge groups (i.e. the $u$-field quanta) and $\beta=1/T$ is the inverse temperature. The formula above is derived for temperatures $T=1/\beta\ll M$ and in a gauge for which $A_0$ is time-independent. The above potential generates a Debye mass for the field $A_0$, the so called ``electric mass''. Notice however that the factor of $N$ in front appears because $N$ excitations (i.e. $N$ components of the field $u$) of the $SU(N)$ fundamental multiplet all interact with the same gauge field, i.e. they carry electric charges and couple to the gauge field as $e^{\pm i A_0\beta}$. This fact is crucial for understanding the phase-transition at zero temperature discussed  by Affleck.

An instanton in this language corresponds to an event in the effective quantum mechanics for which the tunneling $A_0:0\rightarrow 2\pi/\beta$ occurs. 
Such instantons are solutions of the effective quantum action containing \eqref{eq:VA0_therm}, and for this reason they were dubbed \emph{quantum instantons} in \cite{Affleck:1979gy}. Affleck found that their action scales with $N$, rendering $\theta$-dependent part of the vacuum energy exponentially small
\be\label{eq:theta_inst}
E(\theta)\propto-e^{-N(\dots)}\cos\theta \;.
\ee

This is in direct distinction from the calculation done by Witten for zero temperature \cite{Witten:1978bc}
\be\label{eq:witten}
E(\theta)\propto \theta^2/N\;, \theta\in(-\pi,\pi)
\ee
so that the $\theta$-dependence behaves as $E(\theta)\sim N^{-1}$, which is also confirmed on the lattice \cite{Vicari:2008jw}.  
This means that the quantity $N\partial_{\theta}^2E(\theta)\big|_{\theta=0}$ experiences a jump from $0$ at finite temperature to a constant at exactly zero temperature, so that a second order phase transition occurs exactly when zero temperature is reached. 
The different behaviors of the two regimes can be traced to the fact that while the Boltzmann factor of the ``thermal'' excitations is exponentially tiny for the small temperatures $e^{-M\beta}$, the $N$ multiplicity of such excitations necessarily amplifies their contributions if the $N\rightarrow\infty$ limit is taken first. Further, an appearance of the potential $\eqref{eq:VA0_therm}$ is akin to the thermal gluons which cause electric screening and deconfinement \cite{Gross:1980br}. 

Employing the twists $\Omega$ roughly results in a replacement $\beta\rightarrow NL$, which, for $NL\gg M^{-1}$ turns the potential \eqref{eq:VA0_therm} into
\be\label{eq:twist_Veff}
V_{eff}(A_0)\propto -e^{-MLN}\cos(NA_0L)\;.
\ee
The reason for this is that now the only allowed thermal excitations come in combinations of $N$ fundamental $u$-quanta, as discussed below \eqref{eq:Omega_fund}. Now note that the $L\rightarrow \infty$ and $N\rightarrow\infty$ both cause the potential to vanish, so that the two limits commute, eliminating the phase transition at $L=\infty$. If $N$ is large, but finite, the potential for $A_0$ is negligible, and the effective theory is a particle on a circle with a $\theta$ angle for which it is well known that the behavior \eqref{eq:witten} holds. 

So what happened to instantons? Recall that instantons amounted to the spatial transitions $A_0:0\rightarrow 2\pi/L$. However, such tunnelings are not elementary anymore, as the above potential has $N$ inequivalent minima at $A_0=\frac{2\pi k}{NL}$, with $k=0,1,\dots, N-1$. The transitions $A_0:\frac{2\pi k}{NL}\rightarrow \frac{2\pi (k+1)}{NL}$ in fact carry \emph{fractional topological charge $Q_{top}=1/N$} and are avatars of the fractional kink-instantons in regime when $NL\ll M^{-1}$ \cite{Tong:2002hi,Eto:2006pg,Eto:2006mz,Bruckmann:2007zh,Brendel:2009mp,Dunne:2012ae,Dunne:2012zk}. Since here they correspond to transitions in the quantum effective action given by \eqref{eq:twist_Veff} we call them \emph{quantum kink-instantons}.  In contradistinction to \cite{Dunne:2012ae,Dunne:2012zk}, the results we presented are in the opposite limit $NL\gg M^{-1}$, but they can be obtained for any value of $NL$ as long as $N$ is large. The point is that whenever $L<\infty$ the effective theory is always equivalent to quantum mechanics at zero temperature with a potential which is an analytic function of $L$. Keeping in mind that when $LN\gtrsim M^{-1}$, the saddle point of $\lambda$ will develop a smooth dependence on $L$, the partition function will be an analytic function of $L$, confirming the continuity conjectured of Dunne and \"Unsal \cite{Dunne:2012ae,Dunne:2012zk}. 

In fact the quantum mechanics obtained for $NL\gg M^{-1}$ is remarkably similar to the regime of Dunne and \"Unsal. The only difference is that the quantum kink-instanton action is proportional to $e^{-NLM}$ and therefore exponentially small. In fact one can take the point of view that because of the small action such tunnelings can be thought of as in a way ``condensing'' to produce the dependence \eqref{eq:witten}. As $NL$ is decreased, the quantum kink-instantons slowly become semiclassical when $NL\sim M^{-1}$, and they smoothly change into the Dunne-\"Unsal regime $NL\ll M^{-1}$.

\vspace{.5cm}
\noindent{\bf Conclusion:}
We have argued on general grounds that the partition function of theories possessing global $SU(N)$ or $O(N)$ symmetry allows for a construction of the twisted partition function by compactifying the time direction with twisted (as opposed to thermal) boundary conditions. We showed that such twists serve as projectors to the ground state when $N\rightarrow \infty$, for any compactification radius $L$. We then showed explicitly that this is indeed the case for the $CP(N-1)$ and $O(N)$ nonlinear sigma model in any dimension. Finally by specializing to the case of two space-time dimensional $CP(N-1)$ model --- an asymptotically free theory -- we showed that the twisted partition function construction completely eliminates the Affleck phase transition present when twists are removed. The absence of this phase transition was due to the commutativity of the limits $N\rightarrow\infty,\;L\rightarrow\infty$, in contrast to the thermal compactification \cite{Affleck:1979gy}.

Our consideration here not only help explain the apparent success of twisted theories at describing the low energy dynamics in a weakly coupled regime \cite{Dunne:2012ae,Cherman:2013yfa,Dunne:2015ywa}, and give insight into the EK reduction, but are also useful for numerical lattice simulations, significantly reducing the finite volume effects. 
In particular only the states in the $k$-index representation with $k\bmod N=0$ contribute to the partition function, rendering most excitations either exponentially suppressed with $NL$ or linearly suppressed as $L/N$. Further, although we have not discussed the gauge theories, we suspect that considerations similar to what we presented here may be responsible for the apparent success of the improved Twisted Eguchi-Kawai reduction found on the lattice \cite{GonzalezArroyo:2010ss}. 
In the case of pure Yang-Mills theory the only global symmetries are $\mathbb Z_N$ center symmetries. However, in modern terminology, such symmetries are referred to as 1-form symmetries \cite{Gaiotto:2014kfa}, as they act on line objects (i.e. Polyakov loops) and not on point objects. More clearly, such symmetries imply the conservation of fluxes \cite{'tHooft:1979uj}, not ordinary charges, while the excitations are not only particle-like, but can also be string-like. Nevertheless it is compelling to think that similar conspiracies take place as the ones we discussed here. 
In fact twists employed in \cite{GonzalezArroyo:1982ub,GonzalezArroyo:1982hz,GonzalezArroyo:2010ss} are related to the 't Hooft $\mathbb Z_N$ fluxes \cite{'tHooft:1979uj}, which are conserved charges of the center symmetry.
%%%%%%%%%%%%%%%%%%%%%%%%%%%%%%%%%%%%%%%%%%%%%%%%%%%%%%%%%%%%%%%%%%%%%%%%%%%%%

%%%%%%%%%%%%%%%%%%%%%%%%%%%%%%%%%%%%%%%%%%%%%%%%%%%%%%%%%%%%%%%%%%%%%%%%%%%%%

\vspace{0.5cm}
%%%%%%%%%%%%%%%%%%%%%%%%%%%%%%%%%%%%%%%%%%%%%%%%%%%%%%%%%%%%%%%%%%%%%%%%%%%%%%
\begin{acknowledgments}
{\noindent\bf Acknowledgments: }I am grateful to Mohamed Anber, Aleksey Cherman, Gerald Dunne, Erich Poppitz for their useful comments on the draft. I would especially like to thank Mithat \"Unsal for many discussions and comments, especially regarding Twisted Eguchi-Kawai. I would also like to thank Giovanni Villadoro for discussions and hospitality at ICTP. This work has been supported in part by the DOE grant No. DE-FG02-03ER41260.
\end{acknowledgments}
%%%%%%%%%%%%%%%%%%%%%%%%%%%%%%%%%%%%%%%%%%%%%%%%%%%%%%%%%%%%%%%%%%%%%%%%%%%%%%

%\begin{thebibliography}
%\bibliographystyle{apsrev4-1}
\bibliographystyle{JHEP}
%\bibliography{bib_short}
\bibliography{bibliography}
%\bibliographystyle{JHEP}
%\end{thebibliography}

\end{document}